# Examining marginal sequence similarities between bacterial type III secretion system and *Trypanosoma cruzi* surface proteins: Horizontal gene transfer or convergent evolution?

Danielle C. F. Silva[1,2], Richard C. Silva[1], Renata C. Ferreira[2,3] and Marcelo R. S. Briones[1,2]

1 Departamento de Microbiologia, Imunologia e Parasitologia, Universidade Federal de São Paulo, Rua Botucatu 862, ECB 3º andar, CEP 04023-062, São Paulo, SP, Brazil

2 Laboratório de Genômica Evolutiva e Biocomplexidade, Universidade Federal de São Paulo, Rua Pedro de Toledo, 669, 4º andar, CEP 04030-042, São Paulo, SP, Brazil

3 Departamento de Medicina, Disciplina de Infectologia, Universidade Federal de São Paulo, Rua Botucatu 740, CEP 04023-900, São Paulo, SP, Brazil

**Corresponding author:**
Marcelo R.S. Briones
Departamento de Microbiologia, Imunologia e Parasitologia,
Universidade Federal de São Paulo,
Rua Pedro de Toledo 669 4º andar, São Paulo, SP, CEP 04023-062, Brazil.
Tel: 55 11 5576-4537, FAX: 55 11 5572-4711, E-mail: marcelo.briones@unifesp.br

**Abstract**

The cell invasion mechanism of *Trypanosoma cruzi* has similarities with some intracellular bacterial taxa especially regarding calcium mobilization. This mechanism is not observed in other trypanosomatids, suggesting that the molecules involved in this type of cell invasion were a product of (1) acquired by horizontal gene transfer; (2) secondary loss in the other trypanosomatid lineages of the mechanism inherited since the bifurcation Bacteria-Neomura (1.9 billion to 900 million years ago) or (3) *de novo* evolution from non-homologous proteins via convergent evolution. Similar to *T. cruzi*, several bacterial genera require increased host cell cytosolic calcium for intracellular invasion. Among intracellular bacteria, the mechanism of host cell invasion of genus *Salmonella* is the most similar to *T. cruzi*. The invasion of *Salmonella* occurs by contact with the host's cell surface and is mediated by the type III secretion system (T3SS) that promotes the contact-dependent translocation of effector proteins directly into host's cell cytoplasm. Here we provide evidence of distant sequence similarities and structurally conserved domains between *T. cruzi* and *Salmonella spp* T3SS proteins. Exhaustive database searches were directed to a wide range of intracellular bacteria and trypanosomatids, exploring sequence patterns for comparison of structural similarities and Bayesian phylogenies. Based on our data we hypothesize that *T. cruzi* acquired genes for calcium mobilization mediated invasion by ancient horizontal gene transfer from ancestral *Salmonella* lineages.

# 1. Introduction

The protist *Trypanosoma cruzi* is a heteroxenic parasite and the causative agent of Chagas disease which represents an important public health problem in Latin America [1]. Differently from other mammal infecting trypanosomatids, only T. cruzi can actively invade non-phagocitic host cells [2–4]. The cellular invasion mechanism of *T. cruzi* is remarkably similar to invasion mechanisms found in intracellular bacterial genera such as *Shigella* and *Salmonella*, especially regarding cellular calcium mobilization. Because these mechanisms are not observed in other trypanosomatids [2–6] three possible explanations for the origin of *T. cruzi* calcium dependent invasion mechanism can be conjectured: (1) the acquisition by horizontal gene transfer, (2) secondary loss in non-*T. cruzi* trypanosomatids or (3) parallel or convergent evolution from non-homologous *T. cruzi* surface proteins*.*

The "TriTryps" sequencing genome project revealed bacterial kinase genes such as ribulokinase and galactokinases in *T. cruzi* and *Leishmania major* genome [3], consistent with the idea that these kinases were probably acquired by Horizontal Gene Transfer (HGT) from bacteria to trypanosomatids. Also, the hypothesis of HGT was tested to explain the similarity between T. cruzi trans-sialidases and bacterial sialidases [7]*. As a matter of fact,* Opperdoes and Mitchels propose that the acquisition of a large number of foreign genes from viruses and bacteria was necessary for the evolution of trypanosomatids [8].

Similarly to *T. cruzi*, increased host cell cytosolic calcium is required for intracellular invasion of several bacterial genera. Among intracellular bacteria, the mechanism of host cell invasion of genus *Salmonella* shares the highest similarities with *T. cruzi* [6], [9–15]. The i*nvasion of Salmonella* occurs by contact with the host's cell surface and is mediated by the type III secretion system (T3SS) that promotes the contact-dependent translocation of effector proteins directly into host's cell cytoplasm [13], [15–17].

Here we performed exhaustive database searches directed to a wide range of intracellular bacteria and trypanosomatids, exploring sequence patterns and predicted secondary structures for comparison to detect even distant or marginal similarities between sequences and structures of *T. cruzi* that could be even remotely conserved with bacterial type III secretion systems. These conserved structures could be indicative of horizontal gene transfer or an extreme case of convergent evolution very specific in the *T. cruzi* lineage and completely absent in other trypanosomatids.

## 2. Methods

*2.1. Database mining*

*2.1.1. Searches for genes similar to T. cruzi involved in intracellular bacterial invasion*

Nucleotide sequences of genes encoding proteins SipD, SopB, SopD and SopE2, present in all strains of genus *Salmonella* [16] obtained in GeneDB (http://www.genedb.org/Homepage in September/2009), were used as BLASTN queries [18] in completed intracellular bacterial (facultative or obligate) genome (http://www.genedb.org/Homepage in September/2009). New searches were performed in T. cruzi CL-Brener genome database (http://www.genedb.org/Homepage/Tcruzi in October/2009) using the nucleotide sequences from 57 strains of 11 genera and 28 intracellular bacterial species (including S. typhi) obtained in the former search (Supplementary Table S1).

*2.1.2. Searches for T. cruzi proteins similar to T3SS effector proteins from different bacteria*

Amino acid sequences of proteins SipD, SopB, SopD and SopE2 were submitted to BLASTP [18] in the *T. cruzi* CL-Brener protein database (http://www.genedb.org/Homepage/Tcruzi in September/2009). Only the sequences of proteins whose role in calcium mobilization during *T. cruzi* invasion is currently known were selected [19–21] (Supplementary flowchart S2-A). The amino acid sequences from T3SS proteins of *Escherichia coli* (EHEC O157:H7) str. EDL933, *Salmonella enterica* (serovar Typhi) str. CT18, *Shigella flexneri* (serotype 2a) str. 301, *Pseudomonas aeruginosa* PAO1 and *Yersinia pestis* CO92, downloaded from the Virulence Factors Database

(http://www.mgc.ac.cn/VFs/ in March/2010) were also submitted to BLASTP (http://www.genedb.org/Homepage/Tcruzi in March/2010), being selected only the first fifteen sequences according to their lower E-values. The amino acid consensus sequences of *T. cruzi* proteins retrieved from BLASTP, TcCLB.508221.420, TcCLB.510693.150, TcCLB.511089.90 and TcCLB.506611.20 (from this point forward designated as 420, 150, 90 and 20, respectively) were manually mapped and submitted again to BLASTP in the *T. cruzi* genome database GeneDB (http://www.genedb.org/Homepage/ in March/2010) and TriTrypDB – Esmeraldo-like and Non-Esmeraldo-like (http://tritrypdb.org/tritrypdb in April/2010), being selected only the first fifteen non-redundant sequences according to their lower E-values (Supplementary flowchart S2-B).

*2.1.3. Similarity searches in different protists*

Amino acid sequence of *S. typhi* SipD was used as query in numerous searches with BLASTP in the genome database of *Bodo saltans*, *T. brucei gambiense*, *T. brucei 427*, *T. brucei 927*, *T. congolense*, *T. cruzi*, *T. vivax*, *L. mexicana*, *L major* strain *Friedlin*, *L. braziliensis* and *L. infantum* in GeneDB and TritrypDB (http://www.genedb.org/Homepage/ and http://tritrypdb.org/tritrypdb in March/2011), *Euglena gracilis* (txid3039) and *Paramecium teutraurelia* strain d4-2 (txid412030) (http://blast.ncbi.nlm.nih.gov/Blast.cgi in June/2011). Only the first fifteen non-redundant sequences were selected.

*2.1.4. Similarity searches of trypanosomatids and S. typhi*

Genome sequence of *S. typhi* CT18 (chromosome, plasmid 1 and 2) was downloaded from NCBI (http://www.ncbi.nlm.nih.gov/genomes/lproks.cgi in October/2011) and submitted to BLASTN algorithm in the *L. major* strain Friedlin, *T. brucei* strain 927 and *T. cruzi* strain CL Brener genome databases at GeneDB (http://www.genedb.org/Homepage in November/2011). Sequences encoding ubiquitous proteins such as heat shock and mitochondrial were discarded. Amino acid sequences of proteins SipD, SopB, SopD and SopE2 of *S. typhi* were used as query in BLASTP searches in the genome database from *L. major* strain Friedlin and *T. brucei* strain 927 at GeneDB (http://www.genedb.org/Homepage in May/2012).

*2.2. Protein sequence alignments*

The amino acid sequences were aligned using ClustalX [22]. Pairwise alignments were performed using default settings (matrix: Gonnet 250, gap opening = 10.00 and gap extension = 0.10). Multiple alignments were carried out with the following parameters: pairwise and multiple alignments using gap opening and gap extension = 1.00, being the alignment matrix modified to PAM 350 on the protists and trypanosomatids amino acid alignments. Alignments were manually checked and adjusted using the Seaview4 sequence editor [23].

*2.3. In silico analysis of deduced amino acid sequences*

Secondary structure of proteins 420, 150, 90, 20 and SipD were analyzed using Geneious v5.5 [24] with GOR1 method and idc=3 [25]. Protein domain searches were performed in Pfam database [26]. Sequences were also submitted to prediction servers at CBS (http://www.cbs.dtu.dk/services) for signal peptide (SP), transmembrane domains, function and subcellular localization and Post-translational modifications such as N and O-glycosylation. Prediction of GPI-anchor sites (glycosylphosphatidylinositol) was performed by servers GPI-SOM [27] and PredGPI [28]. The membrane proteins were predicted using Mem Type-2L server [29]. The presence of signal sequence of T3SS effector proteins was predicted at Modlab server [30].

*2.4. Codon usage and GC content analysis*

Codon usage analysis was carried out with nucleotide sequences encoding for *S. typhi* SipD and *T. cruzi* proteins 420, 150, 90, 20 and actin (TcCLB.510573.10) using The Sequence Manipulation Suite [31]. The GC content was analyzed using the same sequences and also with their respective upstream and downstream intergenic regions using Geneious v5.5 [24].

*2.5. Sequence variability*

Sequence variability was measure using Shannon entropy [32] with BioEdit v.7 program [33] for each position of the amino acid alignment from full sequences obtained in loopback searches and alignment with the conserved amino acid blocks used in Bayesian phylogenetic trees. Values obtained in nits were converted to bits by calculating the base 2 log of nit values.

*2.6. Phylogenetic inference*

Phylogenetic trees were generated from amino acid sequence alignments retrieved from BLASTP (Supplementary Table S3) and from alignments generated from database searches of different protists (*B. saltans*, *E. gracilis, L. mexicana*, *L. major*, *L. braziliensis* e *L. infantum*, *P. teutraurelia T. brucei gambiense*, *T. brucei 427, T. brucei 927*, *T. cruzi, T. congolense* and *T. vivax*), using MrBayes v3.1.2 [34]. MCMC algorithm started from a random tree, estimating the amino acids substitution model. Trees were inferred from $1 \times 10^7$ generations sampling a tree in every 100 generation until the standard deviation from split frequencies were under 0.01. The parameters and the trees were summarized by wasting at least 25% of the samples obtained (burnin). The consensus trees were then used to determine the posterior probability values. All phylogenetic trees were formatted with the FigTree v1.3.1 program (http://tree.bio.ed.ac.uk/software/figtree/).

## 3. Results and Discussion

*3.1. Proteins involved in bacterial invasion similar to T. cruzi proteins*

Among all bacterial genera analyzed (Supplementary Table S1), positive BLASTN results were obtained only for genera *Bordetella*, *Chlamydophila* and *Shigella.* These sequences, along with sequences encoding proteins SipD, SopB, SopD e SopE2 of *S. typhi* were used as queries for searches in the *T. cruzi* genome database. A total of 689 open reading frames (ORFs) were retrieved. Sequences whose *in silico* translation included frameshifts and/or unrelated amino acids, were excluded. Only amino acid sequences obtained by BLASTP were used for further analysis.

BLASTP searches were then performed using as queries the amino acid sequences of the *S. typhi* effector proteins SipD, SopB, SopD and SopE2 against the *L. major*, *T. brucei* and *T. cruzi* genome database, yielding 21, 24 and 42 sequences respectively. From these sequences, we performed predictions to determine their possible locations and functions (Supplementary Table S4). We show that the number of *T. cruzi* amino acid sequences potentially involved in the invasion mechanism was superior to other trypanosomatids. Two sequences with the potential to be on the parasite surface were found both in *L. major* and in *T. brucei* (Supplementary Table S4). However, they were not analyzed further because they are classified as hypothetical or pseudogenes and because it is already known that both parasites do not mobilize intracellular calcium during invasion and thus cannot actively invade host cells [2–4]. Prediction analysis of *T. cruzi* BLASTP results output showed that 9 sequences had the potential to be involved in host cell invasion (Supplementary Table S4). Among those, only the putative sequences of mucins and/or mucin associated surface proteins (MASP) (420, 150, 90 and 20) were selected because of their already known involvement with calcium mobilization during *T. cruzi* cell invasion [19–21]. To further increase our chance to detect marginal similarities among proteins associated with these mechanisms, search hits of proteins whose involvement in *T. cruzi* cell invasion has not yet been demonstrated, was discarded (Supplementary flowchart S2-B). Positive database search results were only obtained with protein SipD. This protein is known to increase the level of proteins secreted by the T3SS and plays a crucial role in *Salmonella* host cell invasion. Its absence causes the complete impairment of effector proteins translocation and hinders the invasion process [35]. T. cruzi MASPs and mucins and bacterial SipD are expressed on cell surface even before invasion, although these can also be found in the cytosol and are intimately involved with mechanisms of pathogenicity [20], [21], [35–37]. These data suggest the homology among SipD, MASPs and mucins, and also suggest that their functions in calcium mobilization might be conserved [38].

In an attempt to find proteins similar to MASPs and mucins in other T3SS bacteria and not restrict the analysis to proteins associated with calcium mobilization of genus *Salmonella*, we performed new searches against the *T. cruzi* genome database with amino acid sequences from different bacterial T3SS (Supplementary Table S5). These results revealed a considerable number of MASPs and mucins (Table 1) consistent with possible horizontal transfer of genes encoding T3SS to *T. cruzi,* because the blast search results of MASPs and mucins are not unique to *Salmonella* queries. However because the percentage of MASPs returned by searches with *Salmonella* was significantly higher, sequences from other genera were not analysed further (Table 1). Also, when comparing the invasion mechanisms associated with different T3SS, *Salmonella* shows the highest similarity with *T. cruzi*. Both organisms can invade non-phagocytic cells, use inositol 1,4,5-trisphosphate (IP3) to elevate intracellular calcium and consequently induce cytoskeleton rearrangement and remain inside vacuoles during the first stages of cell invasion [6], [9–15]. Although other bacteria share some of these mechanisms, genus *Salmonella* shares most of the observed features. For example, in *Shigella* whose host cell invasion mechanism is relatively similar to *Salmonella* [13] and has T3SS proteins [39], [40], differ in some aspects to T. cruzi invasion mechanism. It does not exclusively depend on intracellular calcium mobilization to invade host cells and it does not remain in vacuoles during the first stages of invasion [9], [10].

To verify if the marginal sequence similarities between bacteria and *T. cruzi* are specific to genes encoding T3SS proteins, searches using the whole *S. typhi* genome as query were performed against the genome databases from different members of Trypanosomatidae (Table 2). These searches returned a large number of sequences coding for common proteins shared by all classes of eukaryotic organisms such as mitochondrial and heat shock proteins. These searches also returned several genes encoding hypothetical proteins and stage-specific proteins of each parasite (data not shown). However, these genes were not considered as positive hits for possible "trace-homologies" that could be involved with infectivity, because negative results were obtained when predictions for subcellular localization, SP and GPI anchoring were performed with their deduced amino acid sequence (data not shown), suggesting that these putative proteins are possibly not secreted or present on the cell surface. These results are supported by the fact that *T. cruzi* adhesion and invasion does not seem to be simple i.e. involving a single ligand-receptor interaction. Trypomastigotes exploit a huge palette of surface glycoproteins, secreted proteases and agonist signaling to actively manipulate the host cell invasion [6], [12], [20], [21], [41–43]. As expected, searches in the *T. cruzi* genome database using the whole *S. typhi* genome returned several sequences that encode proteins involved in host cell adhesion/invasion such as DGF-1 (Dispersed Gene Family 1) and MASPs [19–21], [44] (Table 2).

*3.2. Amino acid sequence similarities*

The complete amino acid sequences of *S. typhi* SipD and of *T. cruzi* MASPs and mucins (420, 150, 90 and 20) were aligned. As expected, due to the high rate of divergence among sequences, it resulted in few conserved blocks and positions embedded in highly divergent domains (data not shown). However, the mapping of local amino acid residues (local alignment) resulted in an alignment with good quality (pairwise identity, identical sites and similarities above 13, 16 and 29% respectively) (Table 3) showing potential homologous positions (Fig. 1). Alignments often provide important insights into protein functional mechanisms being the pairwise alignment of blocks a better option to perform homology searches [38], [45]. SipD has residues important for *Salmonella* invasion. Although most of functional residues are located at the C-terminal, the portion of N-terminal which aligns with the *T. cruzi* proteins also has important sites, both by decreasing the invasion itself and by involvement with bile salts that suppress the *Salmonella* invasion [47], [48]. Although most of the transferred genes are non-functional in the recipient genome, Woolfit et al. [49] suggest that independently of the direction of the HGT, transferred genes may remain functional. These propositions are supported by different authors that argue that these genes are really important in the adaptation to new niches, to originate novel functions and for virulence [8], [50–52].

*3.3. In silico analysis of protein structure and motifs*

To verify possible homologies ("trace-homologies") between *T. cruzi* and *Salmonella* proteins and also address the possible structural and functional properties shared by them, amino acid sequences were analyzed by different prediction methods. Searches for known sequence motifs and domains from manually curated databases using the amino acid sequences of proteins 420, 150, 90 and 20 from *T. cruzi* and the sequence of *S. typhi* SipD, showed that no characterized domains or motifs are present (data not shown). However, our predictions showed that SipD is part of the IpaD family, effector proteins from *Shigella* that share similar functional roles with SipD [39], [40].

As expected, SipD does not present a canonical SP because proteins from the T3SS are secreted through a sec-independent mechanism [53]. The proteins 420, 150, 90 and 20 from *T. cruzi* present potential cleavage sites in positions 21 and 22, 25 and 26, 26 and 27, and 24 and 25 respectively. More importantly, the fact that the possible signal sequences in these proteins remain outside amino acid blocks that aligns with SipD (Fig. 2) suggests that these residues are not cleaved during secretion. Predictions also suggest that proteins 420 and 90 possess possible transmembrane helices between positions 7 and 29, overlapping with their signal sequences. According to Bendtsen et al. [54], transmembrane helices must be disregarded in these cases because signal sequences interfere with these predictions, leading to false positives. In addition, it is known that MASPs are GPI-anchored [20], [43] and that GPI-anchored proteins lack the transmembrane domains [55].

We also found potential GPI anchoring sites in *T. cruzi* proteins 420, 150, 90 and 20 in positions 291, 305, 306 and 145 respectively. As a negative control, the amino acid sequence of SipD was used in this prediction. These data confirm our results because it is already known that MASPs and mucins are GPI-anchored proteins [20], [43]. The potential GPI anchor sites of putative MASPs 420, 150 and 90 are localized at the end of the amino acid sequences that align with SipD. On the other hand, the predicted GPI-anchor site of putative mucin 20 differs from other proteins (Fig. 2), suggesting a potential specialized and/or functional role of this specific site in these MASPs, and supporting their involvement with host-parasite interactions [55-57].

In addition to the comparative results obtained with SipD, putative post translational modifications were analyzed (Table 4). Not surprisingly, the predictions are consistent with already known characteristics of this protein class [20], [43], [58].

The comparison of protein structures is important to reveal evolutionary relationships among proteins. Protein families tend to be structurally conserved and these structures may be maintained even when sequences have diverged beyond any recognizable similarity [59–61]. To verify if the putative *T. cruzi* proteins and *S. typhi* SipD possess conserved secondary structural domains, their local amino acid sequences were analyzed. These local conserved residues are, in general, rare in regions containing sequences of amino acids forming beta-sheets and rich in alpha-helices and coil structures (Fig. 3). The secondary structure of SipD maintains a similarity of approximately 30 to 45% with *T. cruzi* proteins (Table 5). Considering the phylogenetic distance between these organisms, it is reasonable to propose that these levels of secondary structure similarities might indicate homology. However, the quantification of secondary structure predictions should be taken carefully because the current software works with a confidence level of approximately 70% [25], [61], [62]. Nevertheless, our data indicate that the secondary structures of the conserved amino acid regions of *T. cruzi* and *S. typhi* are more conserved than the primary structure (Table 5), mostly because the secondary structure can be maintained even in regions where amino acids are not identical, *via* conservative amino acid substitutions.

*3.4. Horizontal gene transfer and invasion mechanisms*

Although HGT is recognized as an important evolutionary mechanism, its impact has been neglected and confused with mere phylogenetic noise in favor of a vertical signal resulting from the transmission of information from ancestors to descendants [63].

In view of the amino acid similarities and functional analogies between *S. typhi* and *T. cruzi* proteins here presented and because *T. cruzi* is the only trypanosomatid that can actively invade host cells [2–6], we propose the hypothesis of ancient HGT for the origin of calcium dependent invasion of *T. cruzi*. It can be speculated that these ancient HGT events might have occurred by: (1) the ingestion of blood contaminated with *Salmonella spp.* or some other T3SS intracellular bacteria by species of *Triatominae* and the insertion of bacterial genes into the *T. cruzi* genome or (2) insertions and/or gene exchange by endosymbiotic bacteria. We also do not exclude that other trypanosomatids lost their ability to invade since the Bacteria-Neomura bifurcation (secondary loss). Nevertheless, the occurrence of multiple HGT events from bacterial endosymbionts in plants to trypanosomatids described by Hannaert et al. [64] and the possible HGT in trypanosomatids originated from bacteria present in the intestine of *Triatominae* supports our conjecture [8].

Here we examine three possibilities of HGT, summarized in two different "TriTryps" trees, monophyletic (Fig. 4A) and paraphyletic (Fig. 4B). Although most studies agree with the monophyly of genus *Trypanosoma*, this is still a matter of debate [65], [66]. The first HGT hypothesis assumes that the introgression might have occurred at point 1, where T3SS genes were transferred to the ancestor to all "TriTryps". If this is true then all trypanosomatids would carry genes involved in calcium-dependent host cell invasion, but during evolution these genes could have been lost or silenced in all other lineages except for T. cruzi. If HGT occurred at the point 2, T3SS genes would be present only in T. cruzi and T. brucei spp. (Fig. 4A) or if we consider the "TriTryps" tree in Fig. 4B, T3SS genes would be present only in T. cruzi and Leishmania spp. Finally, if the transfer occurred at the point 3, only *T. cruzi* would have acquired the genes to actively invade host cell. Among these three hypotheses, we believe that the third has the highest likelihood due to the relative similarity of the host cell invasion mechanisms of bacteria, such as *Salmonella,* and *T. cruzi* [6], [9–15] and absence of even remotely similar sequences in T. brucei and Leishmania. In addition, this is the most parsimonious hyphothesis because it involves only one acquisition whereas the other hypotheses involve one acquisition and one or two secondary losses (Fig. 4). This hypothesis is also supported by computational predictions (Supplementary Table S4), by the highly superior number of sequences obtained in database searches in *T. cruzi* genome database and by the potential of these sequences to be involved in invasion mechanisms. Although in small numbers, searches against the genome of *L. major* and *T. brucei* returned 2 amino acid sequences. This may suggest that HGT occurred in a trypanosomatid common ancestor and that all other trypanosomatids have lost this mechanism. The vertical inheritance would imply secondary losses dating to the bifurcation Bacteria-Neomura between 1.9 billion and 900 million years ago (Proterozoic Eon), while convergent evolution would imply a *de novo* independent origin of calcium mediated invasion in Bacteria and *T. cruzi* [67].

There are different ways to detect patterns and signs of HGT events. In general these are based on bio-computational analysis, including homology searches, codon usage, GC content analysis and phylogenetic inference [68], [69]. These methods commonly search for the distribution of atypical genes in different organisms and may include the identification of: (a) genes with highly restricted distributions, present in isolated taxa but absent from closely related species, (b) highly similar genes and (c) genes whose phylogenies are incongruent with the relationships inferred from other genes in their respective genomes [70]. Nonetheless, most methods used to evidence HGT are based on recent events, since ancient HGT events are harder to detect and genes may lose ancestor signatures through evolution. Phylogenetic inference of a broad range of sequences, however, may reveal ancient horizontal gene transfers [72], being considered gold-standard.

Parametric analyses such as codon usage and GC content profiles are preferentially used to detect recent HGT events [72]. We analyzed the codon usage profiles of nucleotide sequences encoding the putative *T. cruzi* proteins and *Salmonella* SipD. These analyses were performed with the four-fold degenerated amino acids only. These results did not strongly indicate the occurrence of HGT, but it is noticeable that the codon usage pattern of actin differ from other *T. cruzi* genes (Fig. 5), suggesting a possible HGT event. Although SipD has a different codon usage profile in comparison to *T. cruzi* genes, this cannot be considered a negative result, since highly divergent genes tend to lose features from their ancestors [71], [73]. Additionally, transferred genes tend to behave homogeneously, similar to genes from the receptor organism. Thus, codon usage analyses are not sensitive enough to distinguish ancient HGT [73], [74]. Therefore, if we look carefully it is possible to note that the frequencies of G and C levels in third codon positions are relatively close among genes encoding the *T. cruzi* proteins 420, 150, 90 and 20 and *S. typhi* SipD, in comparison to values of *T. cruzi* actin gene, mainly for the amino acids alanine (ALA), proline (PRO) and threonine (THR) (Fig. 5). Usually vertically inherited genes are adapted to the codon usage characteristic of their original genome and expression level. On the other hand, horizontally acquired genes frequently have atypical G and C base compositions [75]. Together these results support the hypothesis that these *T. cruzi* genes were acquired by horizontal gene transfer, because they have different sequence features when compared to the actin gene.

Selective advantages and gene fixation in the receptor organism are necessary for the compatibility of GC content and codon usage between receptor and donor organisms [76], which is observed in *T. cruzi* and *S. typhi* genes analyzed in this study *as both present* *approximately 51% GC content* [57], [77]. However, most methods identify horizontally transferred genes based on the identification of atypical GC content in DNA sequences [73], [75]. Atypical GC content in intergenic regions may reveal horizontally transferred genome islands [78]. Our results demonstrated that some values were close to the GC content of intergenic and coding regions of each gene, except for the intergenic regions of actin (Table 6). It is known that MASPs and mucins, as well as some other surface proteins, unique to *T. cruzi*, are encoded by non-syntenic islands [3]. Although we have not observed atypical GC content in intergenic regions between the possible genes acquired by horizontal transfer, we do not consider this as a negative result for a possible HGT event, particularly because methods to identify atypical sequences are limited to detection of recent transfers [70] and also because intergenic regions showed lower GC content than the other regions (Table 6). *G*ene content varies along a genome, and the number of members in each gene family. The difference in gene repertoire between the genomes of the same family and/or species is generally attributed to gene loss or HGT [79]. Thus, we can assume that *T. cruzi* may have acquired a large number of foreign genes, because the size of its genome exceeds by 20 Mb the genomes of *T. brucei* and *L. major*, and MASPs and mucins are encoded within large genomic islands [3].

Although entropy analysis is not commonly employed to study possible HGT, here we used Shannon Information Entropy because HGT per se, can be considered a source of disorder. Gene exchange among organisms, populations and species causes extensive genome perturbations and increased mutation [80]. Functional protein sequences are usually more conserved, therefore less entropic, than non-functional sequences [81]. We were then expecting to find lower entropy within the conserved amino blocks in comparison to sequences outside these blocks. Total uncertainty is defined as the maximum number of different amino acids found at the same position. In this case we have 21 different characters (20 different amino acids and gaps), found in the 4 alignments obtained with the sequences from loopback searches. The maximum entropy value in this case is approximately 4.3 bits. Thus, positions with entropy values higher than 2.0 were considered as variable, while positions lower than 2.0 were considered as conserved [44]. In general, our data shows that these aligned amino acid blocks are relatively well conserved, as indicated by the low entropy values and that naturally, gaps have entropy values practically or completely null (Fig. 6).

To obtain a congruent analysis that could establish evolutionary relationships between *S. typhi* SipD and putative *T. cruzi* MASPs and one mucin, a larger number of amino acid sequences were obtained [82] by performing new searches *in the T. cruzi* genome database, using the conserved amino acid blocks from proteins 420, 150, 90 and 20 as queries. This *method reduces* the false positives and increases the chance to find new sequences that could not be discovered by searches with the primary query. The amino acid sequences (Supplementary Table S3) and sequences obtained from database searches of different protists were aligned and submitted to Bayesian phylogenetic inferences. A total of six multiple alignments were generated (one for each *T. cruzi* proteins), comprising up to 36 sequences which included the *S. typhi* SipD, with up to 152 positions, and other 2 alignments, one comprising 170 sequences with 368 positions (different protists) and the other with 139 sequences and 444 positions (only trypanosomatids), obtained by searches in different protein databases. Apart from the phylogenetic inference obtained with the putative mucin 20, which showed a large polytomy (Fig. 7D), all phylogenetic trees inferred with MASPs (420, 150 and 90) *revealed a cluster comprising S. typhi* SipD and several *T. cruzi* proteins, with posterior probabilities above 0.79 (Fig. 7), suggesting a common evolutionary origin. Interestingly, a common feature of trees obtained from the alignments of MASPs 420, 150 and 90 is that few family members of MASPs were closer to SipD than other members within the same family, indicating different subgroups of MASPs with distinct phylogenetic distances relative to SipD. The sequences of putative MASPs of the inference 420 (TcCLB.510693.91 and TcCLB.510693.280) for example, were more divergent in comparison to the rest of MASPs family and forms an outgroup (Fig. 7A). SipD, although more divergent than all the other proteins in the alignments, did not cluster as an outgroup. The MASP (TcCLB.510693.190) that clustered with SipD (Fig. 7A) has been recently described by Santos et al. [83] as MASP16 being highly expressed in bloodstream trypomastigotes and myoblast cells. Therefore, MASP16 as well as other MASPs may be involved in the invasion mechanism and calcium mobilization of *T. cruzi*, suggesting homology and functional analogy of these MASPs with SipD.

The phylogeny inferred using amino acid sequences of different protists was used to test if earlier branching organisms such as *Euglena gracilis*, *Paramecium tetraurelia*, and *Bodo saltans*, as well as other trypanosomatids would cluster together with SipD, which in fact occurred. Although the SipD sequence did not form an exclusive group with *T. cruzi* sequences, it was not absent (Fig. 8). The same results were observed in trees with amino acid sequences of trypanosomatids only. It was observed that SipD is closer to *T. cruzi* with posterior probability 1.00 (Fig. 9). This result may confirm our hypothesis of HGT from intracellular bacteria, more specifically from *Salmonella spp* to *T. cruzi*, because even with a large number of sequences from different trypanosomatids, SipD still clustered together with *T. cruzi* sequences.

The accuracy with which phylogenies can be reconstructed, and by which HGTs can be detected, depends on the degree of divergence [70], [82] and for highly divergent sequences, the number of amino acid substitutions may be saturated, resulting in loss of the phylogenetic signal [46], [70], [74]. Of note, it has recently been shown that the heterologous expression of two *T. cruzi* MASP family member proteins in *Leishmania tarentolae* (non infective to mammal cells) triggers intracellular calcium transients in HeLa cells, presumably by injury to the cell membrane [84]. This observation is consistent with our prediction of functional analogy with *Salmonella* SipD and the horizontal gene transfer here proposed.

## 4. Conclusions

Our results are consistent with the hypothesis that genes involved in host cell invasion were horizontally transferred from *Salmonella-like* ancestors in the early evolutionary history of *T. cruzi.* Because of marginal sequence similarities involved and long divergence dates, our data cannot rule out extreme convergent evolution. Nevertheless, the acquisition of ancestra*l Salmonella* T3SS might have contributed to the pathogenicity and singular invasion mechanisms among trypanosomatids that allowed it to actively invade host cells. We believe that our study presents a provocative hypothesis to stimulate future functional complementation experiments between MASP members and Salmonella type III secretion system proteins, to understand the role of individual proteins of the MASP family in *T. cruzi* invasion.

**Acknowledgments**

We thank Thais F. Bartelli for careful review of the manuscript. DCFS and RCS received fellowships from Coordenação de Aperfeiçoamento de Pessoal de Nível Superior (CAPES), Brazil. This work was supported by grants to MRSB from Fundação de Amparo à Pesquisa do Estado de São Paulo (FAPESP), Brazil; from Conselho Nacional de Desenvolvimento Científico e Tecnológico (CNPq), Brazil and the International Program of the Howard Hughes Medical Institute (HHMI).

**Table 1.** Database searches using amino acid sequences of the T3SS proteins of different bacteria.

| Bacteria | T3SS Proteins | MASP | TcMUCII | Others | MASP (%) |
|---|---|---|---|---|---|
| *E. coli* | 18 | 22 | 8 | 103 | 13.53 |
| *S. typhi* | 8 | 16 | 2 | 50 | **23.53** |
| *S. flexneri* | 6 | 11 | 3 | 61 | 14.66 |
| *P. aeruginosa* | 37 | 23 | 3 | 263 | 7.96 |
| *Y. pestis* | 41 | 20 | 10 | 332 | 5.52 |

**Table 2.** Comparative genome analysis of *S. typhi* and trypanosomatids.

| *S. typhi* | *T. cruzi* | | | *T. brucei* | | | *L. major* | | |
|---|---|---|---|---|---|---|---|---|---|
| | Surface | Hypothetical | Common | Surface | Hypothetical | Common | Surface | Hypothetical | Common |
| Chromosome | 9 (MASPs) | 5 | 86 | 0 | 2 | 98 | 0 | 2 | 99 |
| Plasmid 1 | 97 (DGF-1) | 1 | 2 | 0 | 0 | 4 | 0 | 73 | 31 |
| Plasmid 2 | 3 (MASPs) | 0 | 1 | 0 | 1 | 1 | 0 | 2 | 4 |
| Total | 109 | 6 | 89 | 0 | 3 | 103 | 0 | 77 | 134 |

**Table 3.** Sequence similarities between *Salmonella* SipD and *T. cruzi* MASPs and mucin. Similarity percentages were calculated using Geneious v5.5. software.

| Alignment  | Positions | Identical sites | Pairwise identity | Similarity |
|------------|-----------|-----------------|-------------------|------------|
| SipD X 420 | 145       | 24.8%           | 23.8%             | 37%        |
| SipD X 150 | 142       | 18.3%           | 14.7%             | 30%        |
| SipD X 90  | 142       | 19.7%           | 16.2%             | 32%        |
| SipD X 20  | 88        | 15.9%           | 12.9%             | 29%        |

**Table 4.** Predictions of protein sequence features. The numbers indicate the sites predicted.

| Prediction | SipD | 420 | 150 | 90 | 20 |
|---|---|---|---|---|---|
| Signal peptide | No | Yes | Yes | Yes | Yes |
| Transmembrane helix | No | Yes | No | Yes | No |
| GPI anchors | No | Yes | Yes | Yes | Yes |
| N - Glycosylation | No | 2 | 3 | 3 | 2 |
| O - Glycosylation | No | 32 | 25 | 26 | 38 |

**Table 5.** Comparison of primary and secondary structure similarities. Data were generated from 137 positions respective to SipD.

| Sequences | Primary structure | | | Secondary structure | | | | | Similarity (%) | |
|---|---|---|---|---|---|---|---|---|---|---|
| | Conserved | Identical | Similar | Conserved | α-helix | β sheet | Coil | Turn | Primary | Secondary |
| SipD X 420 | 52 | 34 | 18 | 61 | 40 | 3 | 14 | 4 | 37.96 | 44.53 |
| SipD X 150 | 37 | 20 | 17 | 47 | 38 | 0 | 9 | 0 | 27.01 | 34.31 |
| SipD X 90 | 40 | 22 | 18 | 43 | 33 | 0 | 9 | 1 | 29.20 | 31.39 |
| SipD X 20 | 23 | 11 | 12 | 48 | 35 | 3 | 7 | 3 | 16.79 | 35.79 |

**Table 6.** GC content of *T. cruzi* genes and intergenic regions (IG).

| | GC Content (%) | | |
|---|---|---|---|
| Gene | Coding | IG Upstream | IG Downstream |
| 420 | 50.7 | 52.2 | 52.5 |
| 150 | 52.0 | 50.8 | 58.2 |
| 90 | 51.6 | 52.6 | 54.6 |
| 20 | 55.4 | 55.3 | 48.2 |
| Actin | 51.7 | 32.0 | 36.1 |

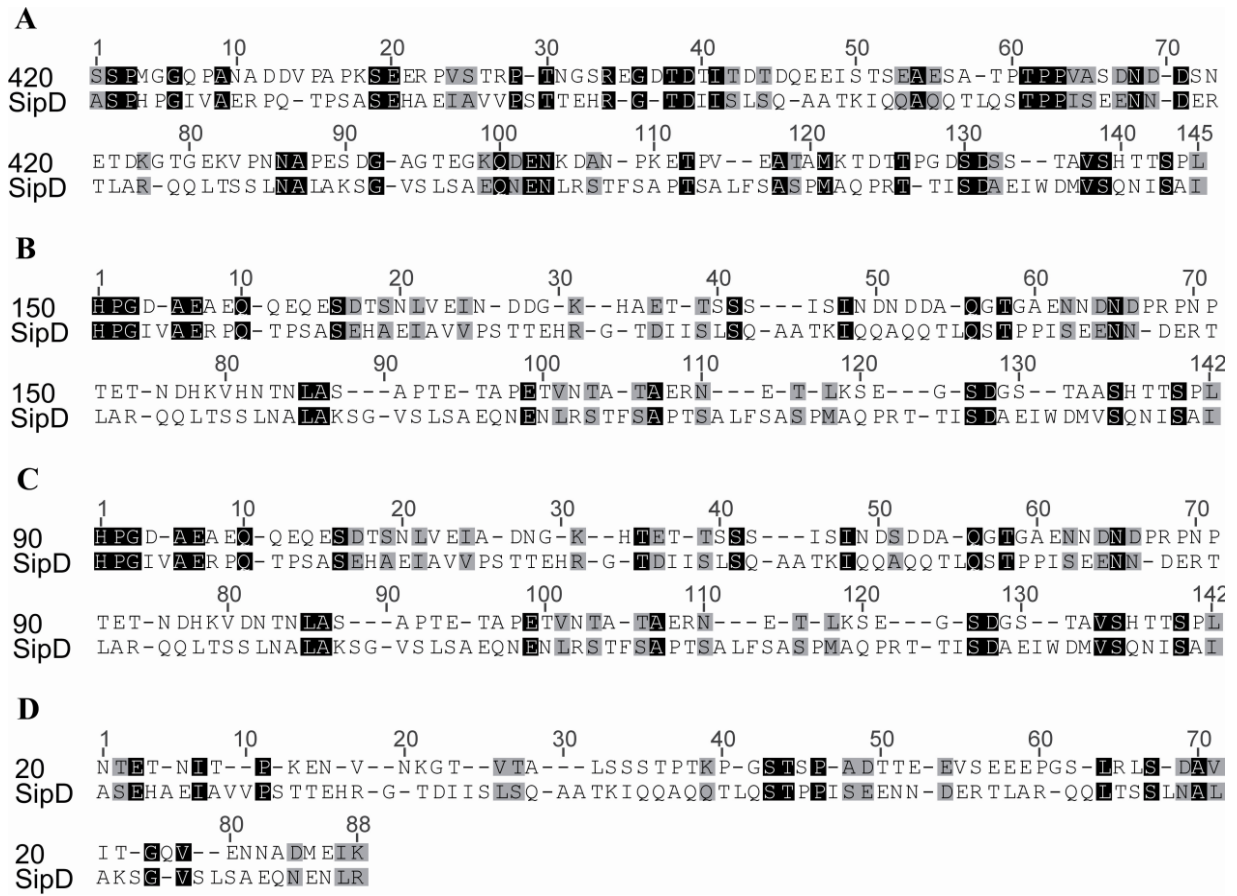

**FIG. 1.** Similarity between *Salmonella typhi* SipD and *T. cruzi* proteins. The identity and similarity between the aligned sequences are represented by shading. Black represents identical residues and gray indicates conservative changes. Local amino acid sequences were initially aligned using ClustalX [22]. Pairwise alignments were performed with default settings (see Methods section) and adjusted manually in Seaview sequence editor [23]. A, B, C and D refer to the local alignment of the amino acid sequence of SipD protein with MASPs 420, 150, 90 and mucin 20 respectively.

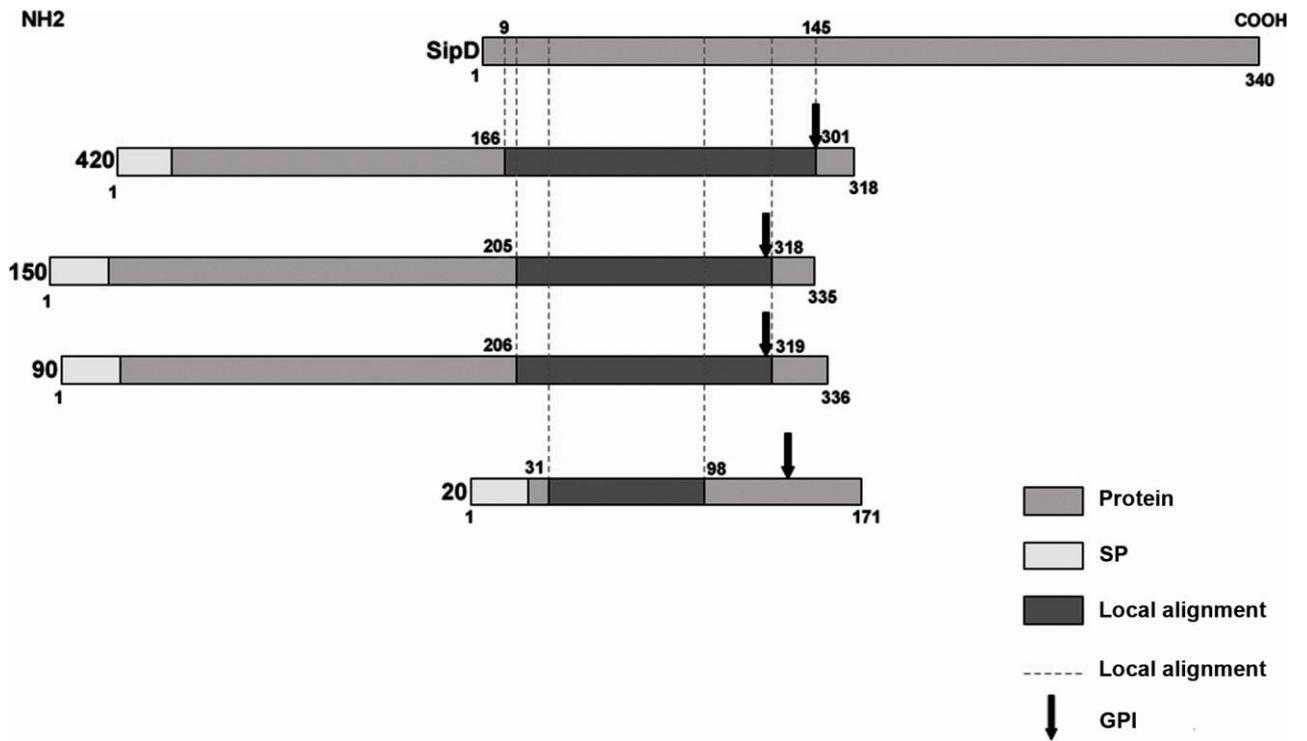

**FIG. 2.** Schematic illustration of amino acid sequence features analyzed by prediction programs.

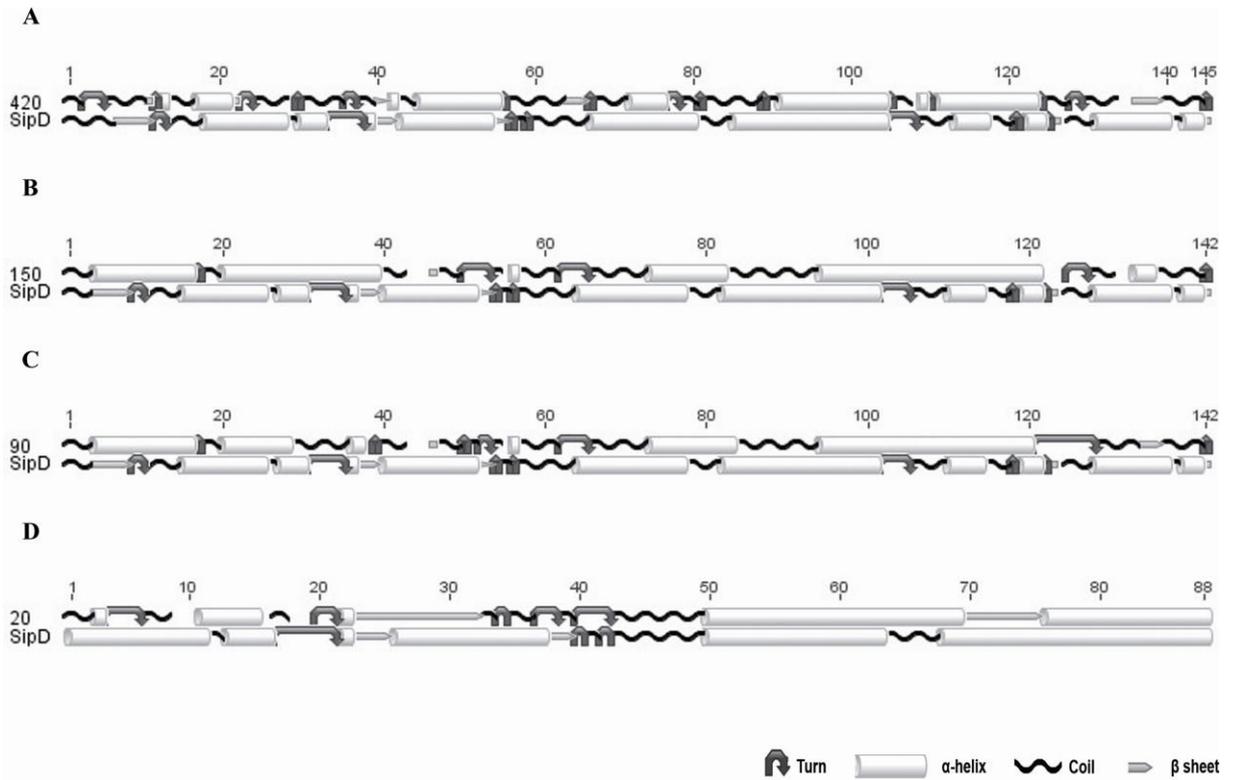

**FIG. 3.** Conserved secondary structure of the aligned blocks of proteins A: 420, B 150, C 90 and D: 20 of *T. cruzi* with SipD. The secondary structure was predicted using Geneious v5.5 software with GOR1 method and idc 3 [24].

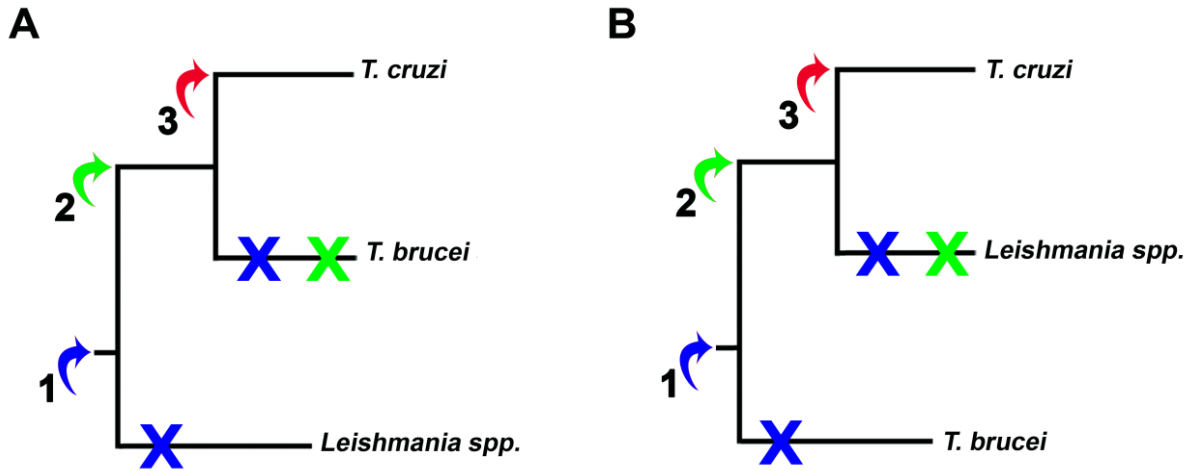

**FIG. 4.** Representation of three different HGT hypotheses respective to "TriTryps" analysed in this study. Arrows indicate the point of HGT and crosses indicate secondary losses. Trees represent the branching order considering monophyly (A) and polyphyly (B) of genus *Trypanosoma*. Hypothesis 1 (blue arrow and crosses) assume very ancient introgression and two secondary losses; hypothesis 2 (green arrow and cross) one introgression and one secondary loss while hypothesis 3 requires only one introgression and no secondary losses. Hypothesis 3 is therefore the most parsimonious because involve a smaller number of events. Also, because of its lower complexity (less assumptions) it can be considered as the null hypothesis while hypotheses 1 and 2 and convergent evolution might be alternative, more complex (more assumptions), hypotheses.

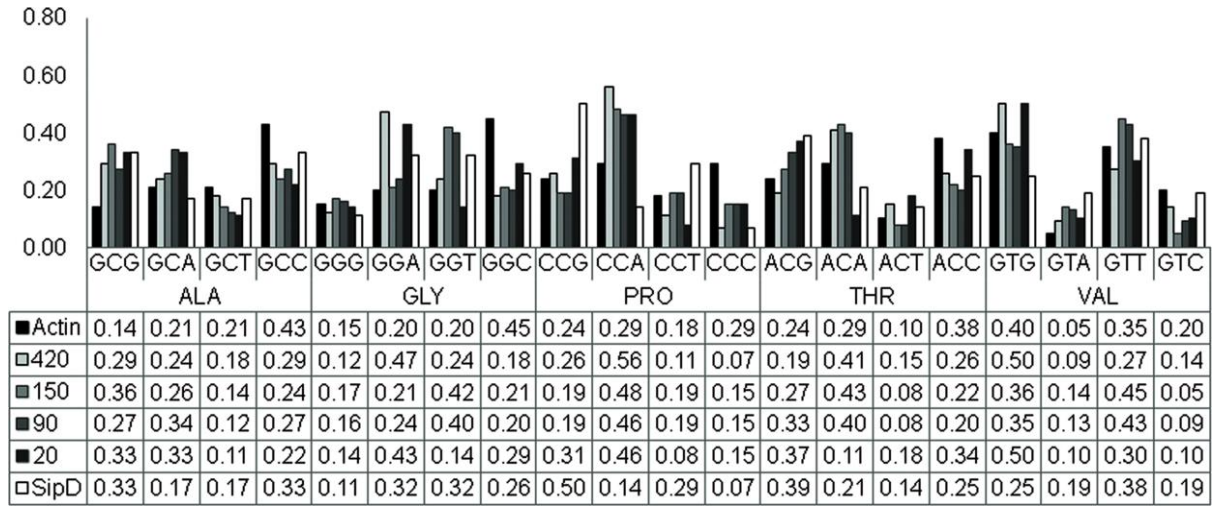

**FIG. 5.** Codon usage profiles. The pattern of codon usage was obtained from the nucleotide sequences coding for proteins SipD, 420, 150, 90, 20 and the actin gene within The Sequence Manipulation Suite [31]. The charts were plotted with the Excel program. The abscissa indicates the four-fold degenerated amino acids and the ordinate represents the codon frequency values. Bars represent each codon used by the respective gene, and the values below the chart indicate the frequency of each codon in the respective genes.

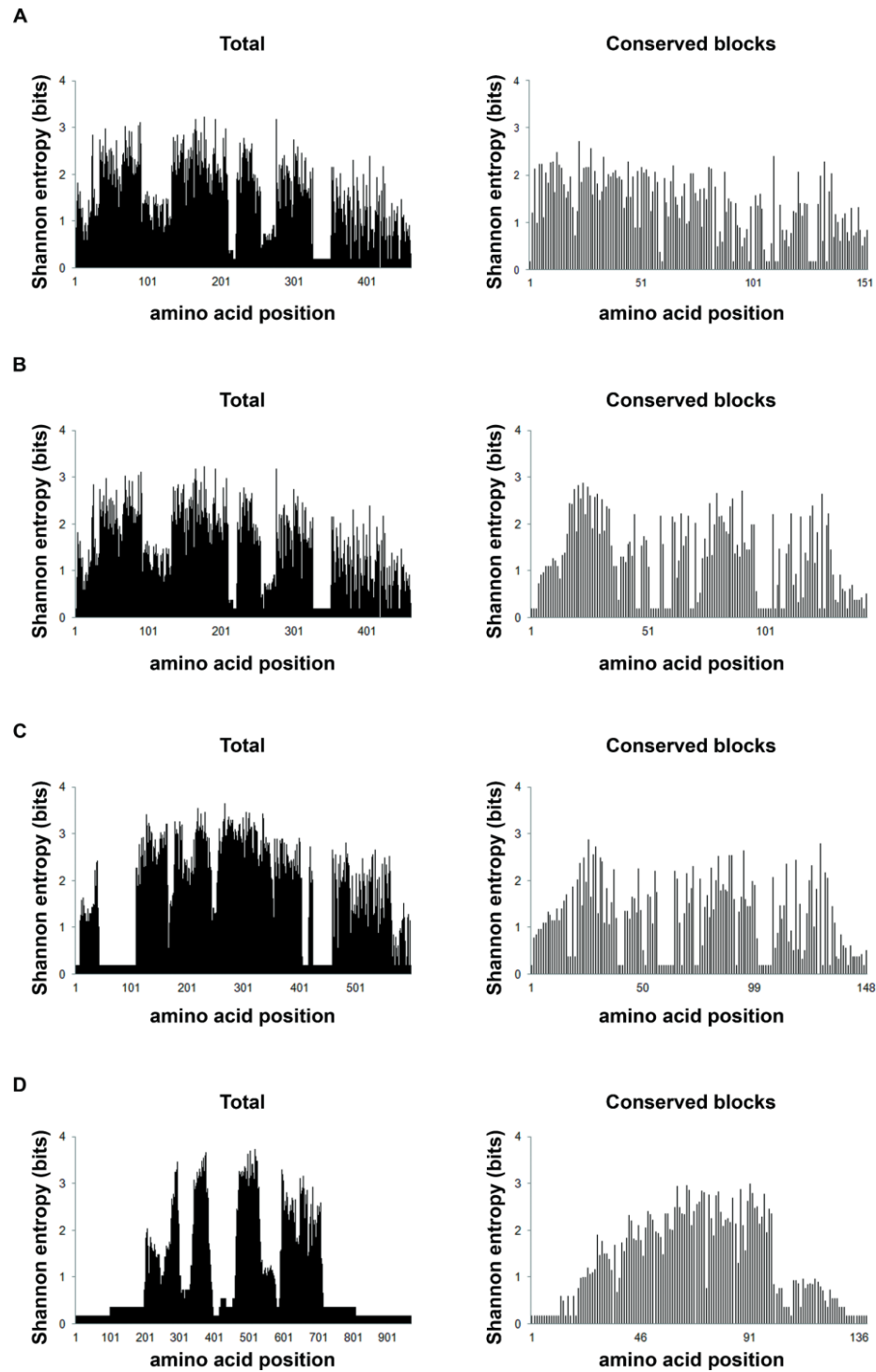

**FIG. 6.** Positional entropy. Shannon information entropy values for the eight different amino acid alignments (full sequences and conserved amino acid blocks) were plotted according to the values generated from BioEdit [33]. The chart A (420), is represented by alignments with 35 sequences, 460 positions (total) and 34 positions (blocks); B (150), by 34 sequences, 460 (total) and 144 (blocks) positions; C (90), 34 sequences, 598 (total) and 148 (blocks) positions; D (20) represented by alignments with 36 sequences and 967 (total) and 139 (blocks) positions. The abscissa represents the positions in each alignment and the ordinate represents the entropy values in bits for each alignment position.

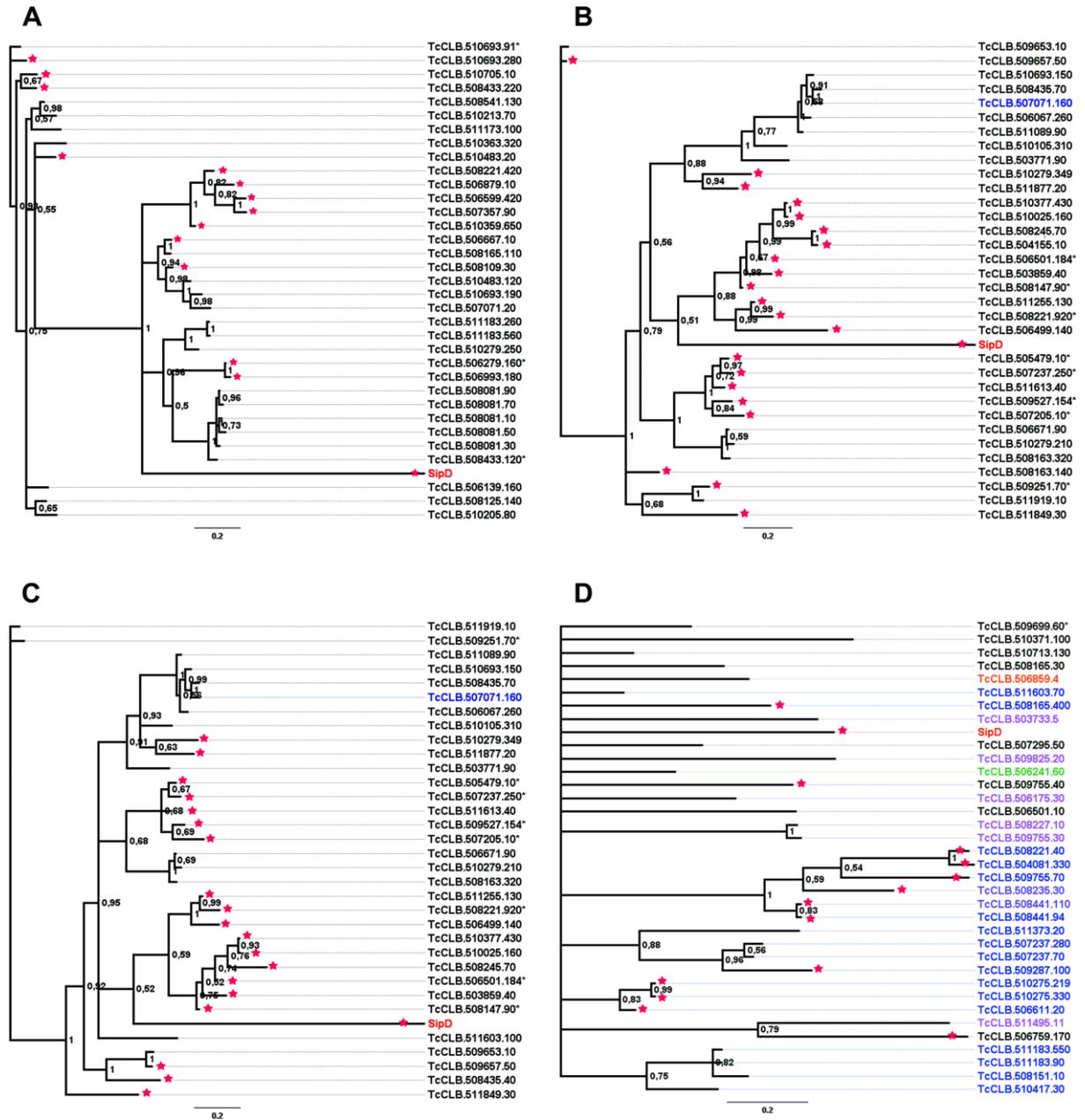

**FIG. 7.** Bayesian phylogeny of MASPs, mucins and *Salmonella* SipD. Trees were inferred with the conserved amino acid blocks obtained by loopback searches. The tree's named 420 (A), 150 (B) and 20 (D) were calculated from $1 \times 10^7$ generations and the tree 90 (C) were calculated from $1.5 \times 10^7$ generations. Numbers in branches represents the posterior probabilities. Letters and numbers on the right side represent GeneDB and TriTrypDB proteins access codes. Different colors indicate the types of proteins, black: MASP, blue: mucins and red: SipD (other colors, check Supplementary Table 4). Asterisks and stars within the codes represent pseudogenes and positive predictions for T3SS proteins respectively.

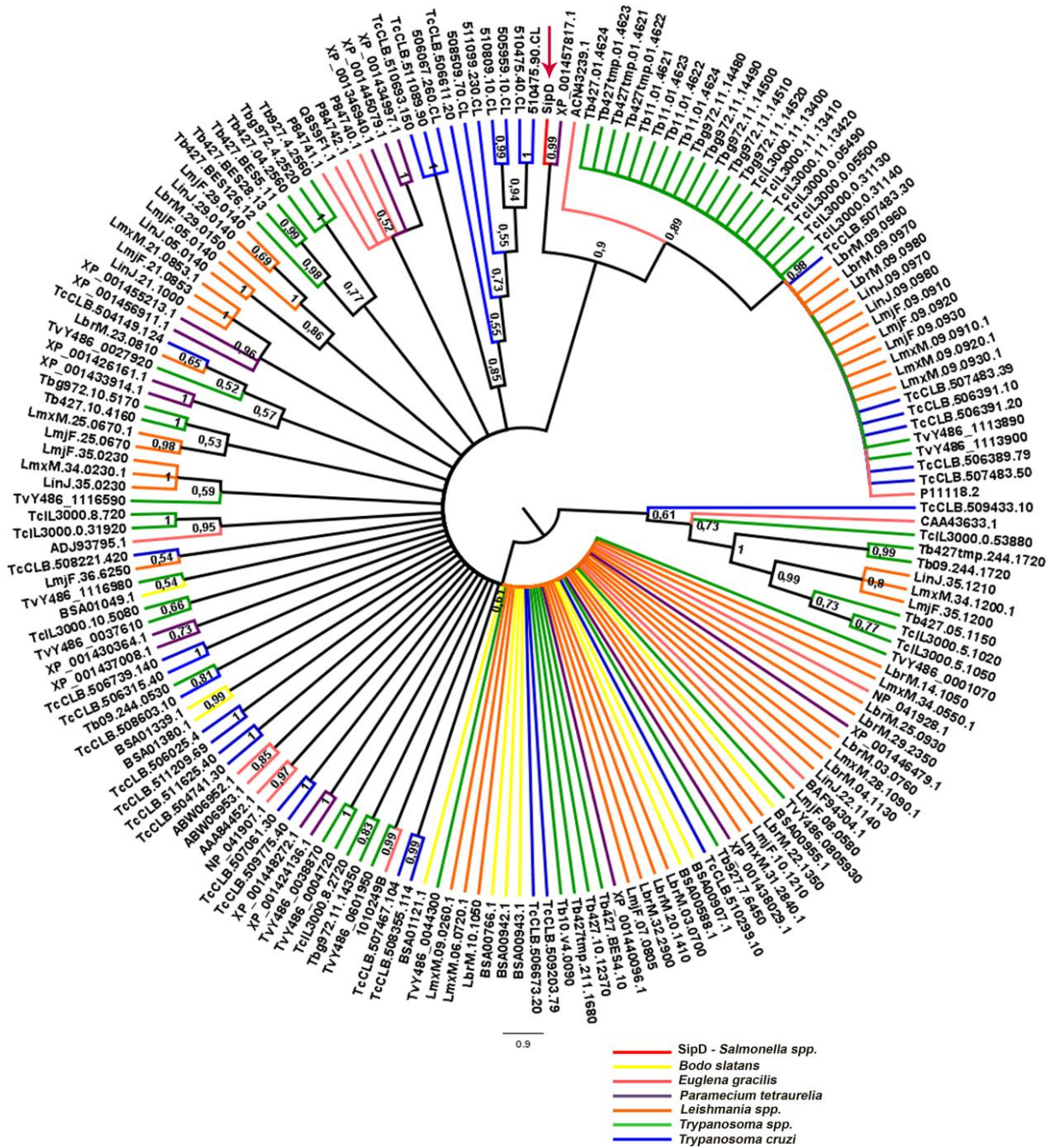

**FIG. 8.** Bayesian phylogeny with different protists and SipD. Trees were inferred with the conserved amino acid blocks obtained by BLASTP of different protists and were calculated from 3 x 10$^7$ generations. Trees are depicted as midpoint rooted. Branches are colored according to genus of protists and numbers in branches represent the posterior probabilities of nodes. Letters and numbers along the branches represent GeneDB, TriTrypDB and NCBI access codes. Arrow indicates the position of the *Salmonella* SipD.

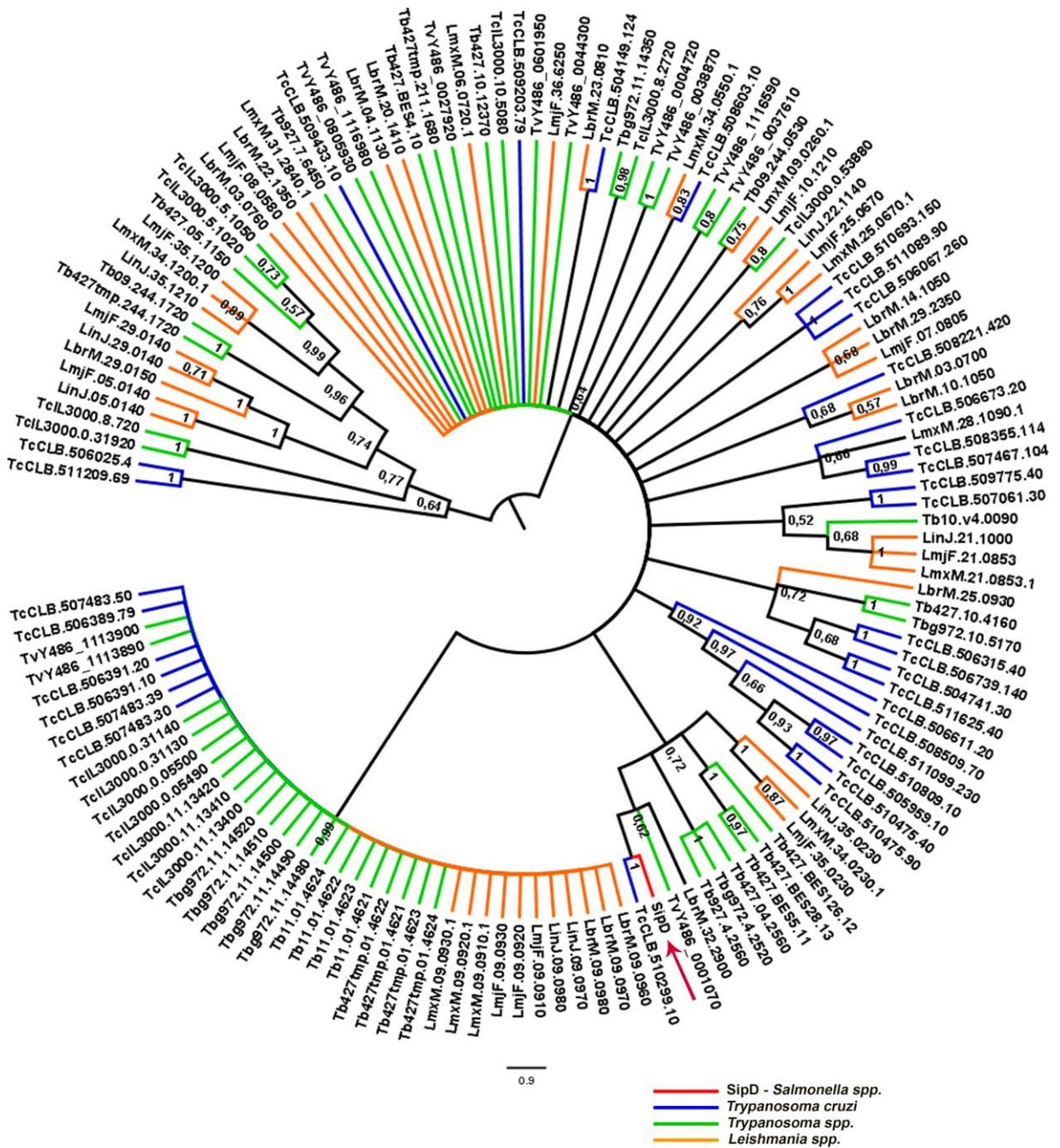

**FIG. 9.** Bayesian phylogeny of trypanosomatids and SipD. Trees were inferred with the conserved amino acid blocks obtained by BLASTP of trypanosmatids and were calculated from $2 \times 10^7$ generations. Trees are depicted as midpoint rooted. Branches are colored according to genera and numbers in branches represent the posterior probabilities of nodes. Letters and numbers along the branches represent GeneDB and TriTrypDB proteins access codes. Arrow indicates the position of the *Salmonella* SipD.**}**